\documentclass[12pt]{article}

\usepackage[margin=1in]{geometry} 
\usepackage{amsmath,amsthm,amssymb}
\usepackage{textcomp}
\usepackage[dvips,final]{graphicx}         		
\usepackage[greek, english]{babel}                			
\usepackage[hang,bf]{caption}              			
\usepackage{rotating}                     		 		
\usepackage{subfigure}
\usepackage{units}
\usepackage{fancyhdr}
\usepackage{lscape}
\usepackage{amsfonts}
\usepackage{float}
\usepackage{graphicx}
\usepackage{longtable}
\usepackage{fancybox}
\usepackage{setspace} 
\usepackage{units} 
\usepackage{pdfpages}
\usepackage{lscape} 		
\usepackage{pdflscape}

\usepackage{upgreek}

\usepackage{listings}

\newcommand{\ssa}{\rule{5mm}{0mm}}
\newcommand{\ssb}[1]{\ssa\mbox{#1}\ssa}
\newcommand{\etal}{\textit{et al. }}

\begin{document}
\bibliographystyle{elsearticle-num}

	\title{Steady state cyclic behaviour of a half-plane contact in partial
		slip subject to varying normal load, moment, shear load, and moderate differential
		bulk tension} 
\author{H. Andresen$^{\,\text{a,}}$\footnote{Corresponding author: \textit{Tel}.: +44 1865 273811; \newline \indent \indent \textit{E-mail address}: hendrik.andresen@eng.ox.ac.uk (H. Andresen).}$\,\,$, D.A. Hills$^{\,\text{a}}$, J.R. Barber$^{\,\text{b}}$, J.V\'azquez$^{\,\text{c}}$\\ \\
	\scriptsize{$^{\text{a}}$ Department of Engineering Science, University of Oxford, Parks Road, OX1 3PJ Oxford, United Kingdom} \\
	$\,$\scriptsize{$^{\text{b}}$ Department of Mechanical Engineering, University of Michigan, Ann Arbor, MI 48109-2125, USA}\\
	$\,$\scriptsize{$^{\text{c}}$ Departamento de Ingenier\'ia Me\'canica y Fabricaci\'on, Universidad de Sevilla, Camino de los Descubrimientos,} \\ 
	\scriptsize{41092 Sevilla, Spain}}

\date{}
\maketitle
	
	\begin{center}
		\line(1,0){470}
	\end{center}
	\begin{abstract}
		\footnotesize{A new solution for a general half-plane contact in the steady state is presented. The contacting bodies are subject to a set of constant loads - normal force, shear force and bulk tension	parallel with the interface - together with an oscillatory set of the same quantities. Partial slip conditions are expected to ensue for a range of these quantities. In addition, the line of action of the normal load component does not necessarily need to pass the centre-line of the contact, thereby introducing a moment and asymmetry in the contact extent. This advancement enables a mapping to be formalised between the normal and tangential problem. An exact and easy to apply recipe is defined.\\}

		\noindent \scriptsize{\textit{Keywords}: Contact mechanics; Half-plane theory; Partial slip; Varying normal and shear loads; Moment; Moderate bulk tension, Mapping}
	\end{abstract}
	\begin{center}
		\line(1,0){470}
	\end{center}

\section{Introduction}

\hspace{0.4cm} We have recently published a partial-slip contact solution \cite{Andresen_2019_2} for the half-plane problem in which a contact is formed by the application of a normal load $P$, and a shear force $Q$ is also gradually exerted while differential tensions $\sigma$ arise in the surfaces of the contacting bodies. These quantities  then vary periodically with time and in phase with each other, so that the trajectory in ($P$, $Q$, $\sigma$) space consists of a line from the origin to some point in the steady state followed by reciprocating behaviour along a straight line
between two points whose separation from the mid-point ($P_0$, $Q_0$, $\sigma_0$) is ($\pm\Delta P/2$, $\Delta Q/2$, $\Delta\sigma/2$). This procedure is very useful for analysing a range of practical problems, such as the gas turbine fan-blade dovetail root contact, and is straightforward to implement, but it lacks one feature present in the prototype. In the dovetail problem, as in other contact problems of this class, the line of action of the normal load does not necessarily pass through the centre-line of the contact, and it generally varies with time. Thus, a moment, $M$, develops and the contact extent becomes asymmetric. 

The purpose of this sequel is to remedy that deficiency, though at the expense of greater algebraic effort. Following
the same pattern of representation as used before, the mean moment developed will be denoted by $M_{0}$ and its range by $\Delta M$. The oscillatory behaviour is in phase with the other changes experienced by the contact and, as before, we neglect the transient problem and argue that, because the prototype experiences many tens of thousands of oscillatory cycles for each major change of load, it is the steady state behaviour which is of most practical importance. We further argue that, when once the permanent stick zone is known, together with the evolution of the contact patch, the most important information needed for a fretting-fatigue analysis is effectively established. We then know the maximum extents of slip, attained just before the points of load reversal are reached, and other things such as the slip displacement can be evaluated afterwards.

\begin{figure}[t]
	\centering
	\includegraphics[scale=0.7, trim= 0 0 0 0, clip]{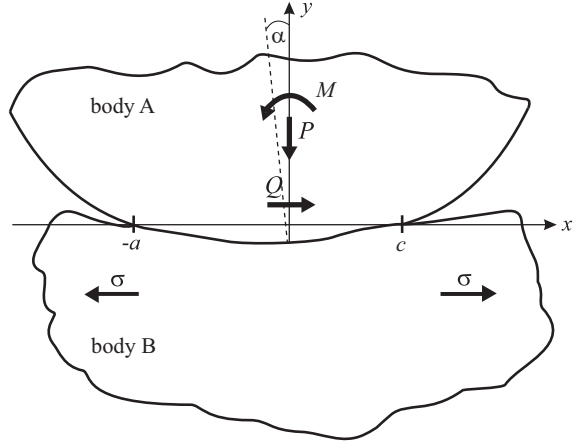}
	\caption{Generic half-plane contact subject to normal load, moment, shear load and bulk tension.}
	\label{fig:generic_half_plane}	
\end{figure}

Figure \ref{fig:generic_half_plane} shows a generic half-plane contact,
subject to normal load, moment, shear load, and bulk tension loading,
for reference. Note, in particular, the sense of a positive applied
moment which will be needed in some of the results to be found.

The majority of partial slip solutions known in the literature employ a method in which the shear traction distribution
is viewed as the sum of that due to sliding, superimposed with a corrective term. 
Hertz was the first to find the solution to the normal contact problem where the contacting bodies
have second order (strictly parabolic but usually interpreted as circular
arc) profiles \cite{Hertz_1881}, so it is natural that the first partial
slip contact solutions were all associated with the same type of geometry. 
The first solution, for a subsequently monotonically increasing shear force, was found by
Cattaneo \cite{Cattaneo_1938}, and, apparently unaware of this solution, Mindlin
\cite{Mindlin_1949} developed the same solution and went on to look at unloading
and reloading problems \cite{Mindlin_1951}, \cite{Mindlin_1953}. These were the only significant
solutions for some time, and then Nowell and Hills \cite{Hills_1987} looked
at what happened when a bulk tension was simultaneously exerted in
one body as the shear force was gradually increased. The
next breakthrough came with the near simultaneous discovery by J\"ager \cite{Jaeger_1998} 
and Ciavarella \cite{Ciavarella_1998} that, just as the `corrective' shear traction was a
scaled form of the sliding shear traction for the Hertz case, the
same geometric similarity applies whatever the form of the contact. Major progress in solving problems involving a varying normal and shear load, and where the intention was to track out the full behaviour
as a function of time, was made in \cite{Barber_2011}. But
this calculation was restricted to $P$-$Q$ problems.

In the analysis which follows it is assumed that the bodies are made from the same material (or more precisely that Dundurs' second constant vanishes) \cite{Hills_1993}. The only restriction on the profile geometry is that it should define an incomplete contact capable of supporting a moment, regardless of whether the geometry itself is unsymmetrical or symmetrical. When an external moment is applied, even a symmetrical `indenter' gives rise to an unsymmetrical contact, and this an important extra feature of the solution absent from this paper's precursor.

\begin{figure}[t]
	\centering
	\includegraphics[scale=0.4, trim= 0 0 0 0, clip]{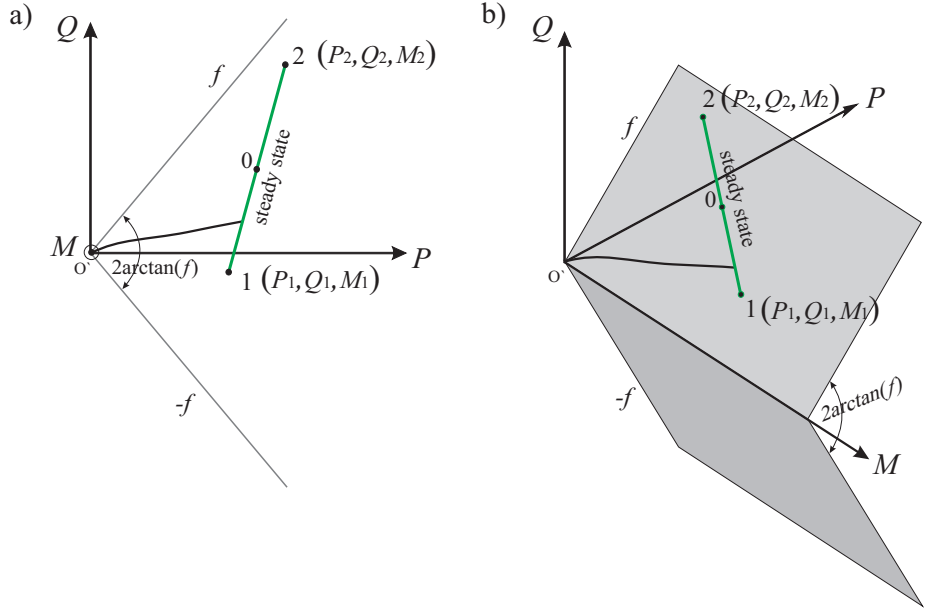}
	\caption{Two (a) and three-dimensional (b) illustration of a load space for a $P$-$Q$-$M$ problem.}
	\label{fig:PMQsigma_load_space}	
\end{figure} 

It is difficult to visualise the problem in an abstract four-dimensional space. Therefore, it is valuable, first, to visualise the history of loading in the two-dimensional load space depicted in Figure \ref{fig:PMQsigma_load_space} (a), ignoring the bulk tension. The initial loading takes us to a point coinciding with the steady state trajectory and the steady-state fluctuations are of range ($\Delta P$, $\Delta Q$, $\Delta M$) so that the loading trajectory moves between points ($P_{1}$, $Q_{1}$, $M_{1}$) and ($P_{2}$, $Q_{2}$, $M_{2}$) passing the mid-point ($P_{0}$, $Q_{0}$, $M_{0}$) every half-cycle. The most likely points to slip are the contact edges, and the contact advances at both edges while going from load point 1 to load point 2.

Note that, while both normal load and moment ($P$, $M$) affect the contact size, and all four
quantities ($P$, $Q$, $M$, $\sigma$) affect the propensity of the contact
to \emph{slip}, only ($P$, $Q$) affect its tendency to \emph{slide}.
In Figure \ref{fig:PMQsigma_load_space} (b) we sketch the
loading trajectory in a three-dimensional load space from a rotated perspective, and
note that the planes implying sliding have a gradient $\pm f$, where
$f$ is the coefficient of friction, with respect to the $P$-axis.
We will assume that the mean and oscillatory load components are such
that the loading trajectory remains, at all times, within the partial
slip `wedge' and note that the solution potentially depends on eight
quantities ($P_{0}$, $Q_{0}$, $M_{0}$, $\sigma_{0}$, $\Delta P$, $\Delta Q$, $\Delta M$, $\Delta\sigma$),
together, of course, with the contact profile, material, and the coefficient of
friction.

\subsection{No slip condition}

\hspace{0.4cm}Before proceeding to the partial slip solution, we start by establishing
the condition for permanent stick, so that no points on the contact
suffer slip at any time. The sign convention shown in Figure \ref{fig:generic_half_plane} is
important in what follows. Suppose that we have a contact whose instantaneous
half-width is $d$, and we make a small change, $\Delta P$, in normal
load together with a small change in moment, $\Delta M$. The corresponding
change in contact pressure is given by \cite{Sackfield_2001}
\begin{equation}
\Delta p(x)=\frac{\Delta P}{\pi\sqrt{d^{2}-x^{2}}}+\frac{2\Delta Mx}{\pi d^{2}\sqrt{d^{2}-x^{2}}}.\label{ec_05}
\end{equation}

If at the same time there are small changes in shear force, $\Delta Q,$
and differential bulk tension, $\Delta\sigma_{0},$ where the latter
is, here, exerted only in body B\footnote{In general, bulk stresses may arise in each body. When this is the
	case, providing they are synchronous, we define $\Delta\sigma=\Delta\sigma_{\text{B}}-\Delta\sigma_{\text{A}}.$}, Figure \ref{fig:generic_half_plane}, the change in shear traction, $\Delta q(x),$ generated is given
by 
\begin{equation}
\Delta q(x)=\frac{\Delta Q}{\pi\sqrt{d^{2}-x^{2}}}+\frac{\Delta\sigma x}{4\sqrt{d^{2}-x^{2}}}\label{ec_02}.
\end{equation}

So, as the most likely points to slip are the contact edges the condition
for no slip if $\Delta P>0$ is 
\begin{equation}
\left(\frac{\Delta Q}{\Delta P}\pm\frac{\pi}{4}\frac{ d\Delta\sigma}{\Delta P}\right)\left/\left(1\mp\frac{2\Delta M}{d\Delta P}\right)\right.<f,\label{ec_06}
\end{equation}
where we choose the upper sign when considering the left hand side (LHS)
of the contact and the lower sign when considering the right hand side (RHS) of
the contact. The necessary condition for no slip is that the inequality
holds at \emph{both} sides of the contact. If this is the case it
follows immediately that, when the loads are reversed and we are re-tracing
our steps on the loading trajectory, full stick will continue to be
maintained, that is the contact will never slip, at any point. Conversely, if the unloading trajectory deviates from loading one, there will be some slip.

\section{Steady state slip behaviour}

\hspace{0.4cm}In this sequel \cite{Andresen_2019_2} we shall continue
to assume that the amount of bulk tension arising is small so that
it is insufficient to reverse the direction of slip at either end
of the contact. 

The slip direction is reversed at each edge when going
from point 1 to point 2 of the loading cycle compared with going from
2 to 1, and the slip zones attain their maximum extent just before
the end points of the loading trajectory are reached. 

\subsection{Normal loading}\label{S-normal}

\hspace{0.4cm} Consider, first, the normal loading problem. At any given point in the loading cycle, the relative slope of the half-plane surfaces, $\mathrm{d}v/\mathrm{d}x$, is given by 
\begin{equation}
\frac{\mathrm{d}v}{\mathrm{d}x}=-\frac{A}{\pi}\int_{-a}^{c}%
\frac{p(\xi)\,\mathrm{d}\xi}{\xi-x}, \label{ec_07}%
\end{equation}
where the contact region is defined as $-a<x<c$ as shown in Figure 1, and the solution to be
developed is mathematically exact only if the materials considered are elastically similar,
having the property $E_A, \nu_A = E_B, \nu_B$, where 
\[
A=4 \left(\frac{(1-\nu^{2})}{E}\right)
\]
 is the material compliance, $E$ being the Young's modulus and $\nu$ Poisson's
ratio  of the bodies (A, B), and plane-strain obtains. If the contacting bodies have a relative
surface normal profile, $g(x)$, and one of the bodies is tilted through
some small angle, $\alpha$, which varies with the applied moment,
the integral equation relating the pressure to the profile is 
\begin{equation}
\frac{\mathrm{d}g}{\mathrm{d}x}+\alpha=\frac{\mathrm{d}v}{\mathrm{d}%
x}=-\frac{A}{\pi}\int_{-a}^{c}%
\frac{p(\xi)\,\mathrm{d}\xi}{\xi-x},\text{ }-a<x<c. \label{ec_08}%
\end{equation}

Normal and rotational equilibrium are imposed by setting
\begin{align}
P  &  =\int_{-a}^{c}p(x)\,\mathrm{d}x,\label{ec_09}\\
M  &  =\int_{-a}^{c}p(x)\,x\,\mathrm{d}x. \label{ec_10}%
\end{align}

In the physical problem, the inputs are the profile, $\mathrm{d}g/\mathrm{d}x$, together with the loads
$P$, $M$, and the outputs are the contact coordinates, $\left[-a, \quad c\right]$, and the angle of tilt, $\alpha$.

\subsubsection{Solution}\label{S-normal-1}

\hspace{0.4cm} The solution of equation (\ref{ec_08}) is
\begin{equation}
p(x)=\frac{w(x,a,c)}{\pi A}  \int_{-a}^{c}\frac{\left(g^\prime(\xi)+\alpha\right)  \mathrm{d}\xi}{w(\xi,a,c)(\xi-x)}\;,\text{ }-a\leq x\leq c \label{ec_12}%
\end{equation}
\cite{Hills_1993}, where $w(x,a,c) = \sqrt{(x+a)(c-x)}$. Equation \eqref{ec_12} is subject to the consistency condition
\begin{equation}
\int_{-a}^{c}\frac{\left( g^\prime(\xi)+\alpha\right)
\mathrm{d}\xi}{w(\xi,a,c)}=0\;, \label{ec_13}%
\end{equation}
and using the identity $\int_{-a}^{c}\frac{\mathrm{d}\xi}{w(\xi,a,c)}=\pi$,
we conclude that
\begin{equation}
\int_{-a}^{c}\frac{g^\prime(\xi)}{w(\xi,a,c)}\mathrm{d}\xi=-\pi\alpha\;. \label{ec_14}%
\end{equation}
Further, the identity 
\begin{equation}
\int_{-a}^{c}\frac{\mathrm{d}\xi}{w(\xi,a,c)(\xi-x)}=0,\text{ }-a\leq x\leq c \label{ec_15}%
\end{equation}
shows that the angle of tilt, $\alpha$, does not explicitly  affect the general
\emph{form} of the inverted integral equation which is given by 
\begin{equation}
p(x)=\frac{w(x,a,c)}{\pi A}\int_{-a}^{c}\frac{g^\prime(\xi)\mathrm{d}\xi}{w(\xi,a,c)(\xi-x)}\;,\text{ }-a\leq
x\leq c\;, \label{ec_16}%
\end{equation}
but merely has an influence on the integration limits, [$-a,\quad c$].

\subsection{Tangential loading}

\begin{figure}[ht]
	\centering
	\includegraphics[scale=0.5, trim= 0 0 0 0, clip]{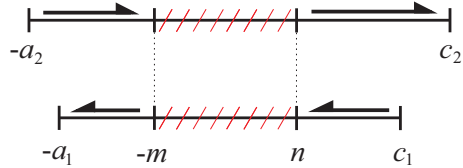}
	\caption{Contact as the ends of loading cycle are approached, including a permanent stick zone.}
	\label{fig:contact_patch}	
\end{figure}

\hspace{0.4cm} We turn now to tangential loading in the steady periodic state. During each half-cycle, the contact starts in a state of complete stick and slip zones grow on each side. We restrict attention to cases in which the direction of slip is the same at either end of the contact during each half-cycle, as shown in Figure \ref{fig:contact_patch}, which represents the extent of the stick and slip zones just before each load reversal.
The stick zone reaches its minimum extent at these points, and this defines the permanent stick zone $\left[-m, \quad n\right]$. Note that $-a_{1}<-m$ and $n<c_{1}$, where $a_{1}$ and $c_{1}$ are the contact coordinates at the point of the loading cycle at which the normal load $P$ is a minimum. Notice also that the stick zones just before each load reversal must be of the same extent, in order for there to be continuity of material over several cycles -- it would not be possible for material to slip out but not to slip back, in the steady state, over any part of the contact.

For load point 1 $(P_1,Q_1,M_1,\sigma_1)$ we write the relative surface strain parallel with the surface, $\Delta\varepsilon_{xx,1}$, as the sum of the effect of the sliding shear traction $-fp_1(x)$ over the whole contact $\left[-a_{1},\quad  c_{1}\right]$ and a corrective term $q_1^*(x)$, at present unknown, over the stick region $\left[-m, \quad n\right]$. Thus,
\begin{equation}
\Delta\varepsilon_{xx,1}=-\frac{A}{\pi}\int_{-a_{1}}^{c_{1}}%
\frac{f p_{1}(\xi)\mathrm{d}\xi}{\xi-x}+\frac{A}{\pi}\int_{-m}^{n}\frac
{q_{1}^{\ast}(\xi)\mathrm{d}\xi}{\xi-x}+\frac{A}{4}\sigma_{1}\;. \label{ec_17}%
\end{equation}

A similar expression can be written for load point 2 $(P_2,Q_2,M_2,\sigma_2)$, except that the sign of the sliding shear traction is
reversed, so as to give the correct sign of shear traction in the end slip regions. We obtain
\begin{equation}
\Delta\varepsilon_{xx,2}=\frac{A}{\pi}\int_{-a_{2}}^{c_{2}}%
\frac{f p_{2}(\xi)\mathrm{d}\xi}{\xi-x}+\frac{A}{\pi}\int_{-m}^{n}\frac
{q_{2}^{\ast}(\xi)\mathrm{d}\xi}{\xi-x}+\frac{A}{4}\sigma_{2}\;. \label{ec_18}%
\end{equation}

In the permanent stick region, the locked in difference between the surface strains must
remain unchanged, i.e. 
\begin{equation}
\Delta\varepsilon_{xx,1}=\Delta\varepsilon_{xx,2}\,\,,\text{
}-m<x<n \label{ec_19}%
\end{equation}
and hence, using equations \eqref{ec_17} and \eqref{ec_18}, 
\begin{equation}
-\frac{A}{\pi}\int_{-a_{1}}^{c_{1}}\frac{f p_{1}(\xi)\mathrm{d}\xi}{\xi
-x}-\frac{A}{\pi}\int_{-a_{2}}^{c_{2}}\frac{f p_{2}(\xi)\mathrm{d}\xi}{\xi
-x}-\frac{A}{4}(\sigma_{2}-\sigma_{1})=\frac{A}{\pi}\int_{-m}^{n}\frac{\left[
q_{2}^{\ast}-q_{1}^{\ast}\right]  \mathrm{d}\xi}{\xi-x}\;,\text{ }-m<x<n\;.
\label{ec_20}%
\end{equation}

Since $-a_{2}<-a_{1}<-m<x<n<c_{1}<c_{2},$ we may make use of the relations
from the normal contact solution, equation \eqref{ec_08}, to rewrite this equation in the
form
\begin{equation}
\frac{2 f}{A}\frac{\mathrm{d}g}{\mathrm{d}x}+\frac{2 f}{A} \alpha_0%
-\frac{\Delta\sigma}{4}=\frac{1}{\pi}\int_{-m}^{n}\frac{\left[  q_{2}^{\ast
}-q_{1}^{\ast}\right]  (\xi)\mathrm{d}\xi}{\xi-x},\text{ }-m<x<n,
\label{ec_21}%
\end{equation}
where 
\[
\alpha_0=\frac{\alpha_{1}+\alpha_{2}}{2}\ssb{and}\Delta\sigma=\sigma_{1}-\sigma_{2}
\]
are respectively the average angle of tilt and the range of differential bulk tension between the two load points. It is worth noting that the LHS of equation (\ref{ec_21}) has just two terms; one is defined by the profile of the contacting bodies (as appears in the normal load problem), and the other is a constant.

\subsubsection{Tangential equilibrium}

\hspace{0.4cm} We denote the resultant \textit{corrective} shear forces by
\begin{equation}
Q_{i}^{\ast}=\int q_{i}^{\ast}(x)\mathrm{d}x\;;\;\;\;i=1,2\;, \label{ec_22}%
\end{equation}
so that we may now impose tangential equilibrium by setting
\begin{equation}
Q_{1}  =-fP_{1}+Q_{1}^{\ast}\;;\;\;\;
Q_{2}  =fP_{2}+Q_{2}^{\ast}\;. \label{ec_24}%
\end{equation}

The range of shear force, $\Delta Q$, is given by
\begin{equation}
\Delta Q=Q_{2}-Q_{1}=2 f P_0+\Delta Q^{\ast}, \label{ec_25}%
\end{equation}
where 
\[
P_0= \frac{P_1 + P_2}{2}
\]
is the mean normal load and 
\begin{equation}
\Delta Q^{\ast}=\int_{-m}^{n}[q_{2}^{\ast}-q_{1}^{\ast}](x)\mathrm{d}x\;.
\label{ec_26}%
\end{equation}

\subsubsection{Solution}

\hspace{0.4cm} The solution of the tangential problem can be facilitated by exploiting significant parallels with the normal contact problem of Section \ref{S-normal}. For example, the inversion of equation
\eqref{ec_21}, bounded at both ends, is given by
\begin{equation}
\lbrack q_{2}^{\ast}-q_{1}^{\ast}](x)=-\frac{w(x,m,n)}{\pi}\int%
_{-m}^{n}\left(  \frac{2f}{A}g^\prime(\xi)+
\frac{2 f}{A} \alpha_0-\frac{\Delta\sigma}{4}\right) \frac{ \mathrm{d}\xi}%
{w(\xi,m,n)(\xi-x)} \label{ec_27}%
\end{equation}
 \cite{Hills_1993}, and arguments exactly parallel to those in Section \ref{S-normal-1} lead to the results
\begin{equation}
\int_{-m}^{n}\frac{g^\prime(\xi)\mathrm{d}\xi}{w(\xi,m,n)}=-\pi \alpha_0 +\frac
{A\pi\Delta\sigma}{8f}\;, \label{ec_29}%
\end{equation}
and
\begin{equation}
\lbrack q_{2}^{\ast}-q_{1}^{\ast}](x)=-\frac{2f w(x,m,n)}{\pi A}%
\int_{-m}^{n}\frac{g^\prime(\xi)\mathrm{d}\xi}{w(\xi,m,n)(\xi-x)}\;,\text{ }-m\leq x\leq n\;. \label{ec_31}%
\end{equation}

This shows that the corrective shear traction is geometrically of
the same form as the contact pressure distribution, but associated
with a particular normal load and moment which will not be the same
as those actually present on the contact.

\subsection{Mapping between the normal and tangential problems}\label{mapping_section}

\hspace{0.4cm} The parallels between the normal and tangential problems identified in Section 2.2.2 can be formalized by defining the mapping
\begin{eqnarray}
[-a, c] &\rightarrow& [-m, n] \nonumber\\
p(x) &\rightarrow& -\frac{1}{2f} \,[q_2^*-q_1^*](x) \nonumber\\
P &\rightarrow& P_0 - \frac{\Delta Q}{2f} \label{mapping}\\
\alpha &\rightarrow& \alpha_0 - \frac{A \Delta\sigma}{8 f}  \nonumber
\end{eqnarray}

If the normal contact problem for an indenter of given profile $g^\prime(x)$ can be solved, meaning that we can determine closed form expressions for the contact coordinates $[-a,c]$ and the pressure distribution $p(x)$ for arbitrary values of $P$ and $\alpha$, the solution to the steady-state tangential problem can then be found readily \textit{mutatis mutandis,} using the above mapping. In many cases, the normal problem will be defined in terms of the forces $P_i$ and the moments $M_i$, $i=1,2$, in which case a preliminary stage will involve the determination of the corresponding tilt angles $\alpha_i$. The procedure is best illustrated by way of a simple example problem, which we present in the next section.

\subsection{Example: the tilted wedge}

\begin{figure}[ht]
 	\centering
 	\includegraphics[scale=0.65, trim= 0 0 0 0, clip]{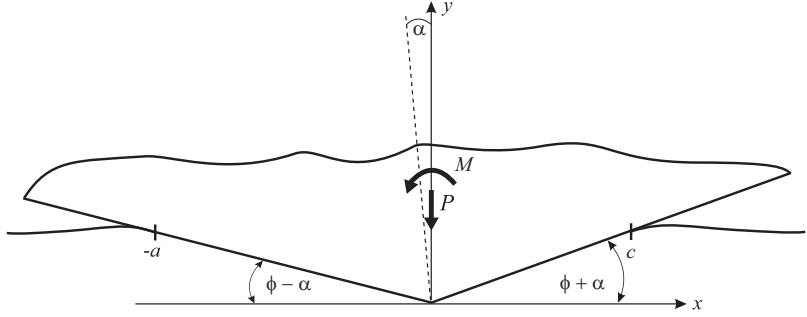}
 	\caption{Tilted wedge subject to a normal load and a moment.}
 	\label{fig:Tiled_wedge}	
 \end{figure}

 \hspace{0.4cm} Figure \ref{fig:Tiled_wedge} shows a shallow wedge of apex angle $(\pi-2\phi)$, $\phi\ll1$, pressed into an elastic half plane by a force $P$ acting through the vertex, and also subject to a moment $M$. The moment causes the wedge to tilt through an angle $\alpha$ as shown. The normal contact problem was solved by Sackfield \etal \cite{Sackfield_2005}, who showed that the pressure distribution is 
 \begin{equation}\label{pressure_distribution_wedge}
 p(x)= \frac{2 \phi}{\pi A} \ln\left\vert \frac{1-\sqrt{(1-x/c)/(1+x/a)}}{1+\sqrt{(1-x/c)/(1+x/a)}} \right\vert\;.
 \end{equation}
 
The contact coordinates are determined from the equilibrium equation
  \begin{equation} \label{normal_equilrbium_wedge}
  P =\int_{-a}^{c}p(x)\,\mathrm{d}x =\frac{2 \phi}{A} \sqrt{a c}\;,
 \end{equation}
and the consistency condition (\ref{ec_13}), which here yields
\begin{equation}\label{consistency_cond_wedge}
 \int_{-a}^{c}\frac{g^\prime(\xi)\mathrm{d}\xi}{w(\xi,a,c) }=-\pi\alpha =2 \phi\, \arcsin\left(\frac{c-a}{a+c} \right)
 \end{equation}
\cite{Sackfield_2005}. These equations can be solved to give
\begin{equation}
a=\frac{AP}{2\phi}\sqrt{\frac{1+s}{1-s}}\;;\;\;\;c=\frac{AP}{2\phi}\sqrt{\frac{1-s}{1+s}}
\ssb{where}s=\sin\left(\frac{\pi\alpha}{2\phi}\right)\;. \label{ca}
\end{equation}

For completeness, we also give the expression for the moment $M$ which is
    \begin{equation}
M =\int_{-a}^{c}p(x) x\,\mathrm{d}x = \frac{\phi}{A} \left(\frac{a+c}{2}\right)^2 \left(\frac{c-a}{a+c} \sqrt{1-\left(\frac{c-a}{a+c}\right)^2} + \arcsin{\left(\frac{c-a}{a+c}\right)} \right)\;.
 \end{equation}

The coordinates $[-m,n]$ defining the permanent stick zone are then immediately determined by substituting the mapping (\ref{mapping}) into (\ref{ca}), giving
\begin{equation}
m=\frac{A}{2\phi}\left(P_0-\frac{\Delta Q}{2f}\right)\sqrt{\frac{1+t}{1-t}}\;;\;\;\;n=\frac{A}{2\phi}\left(P_0-\frac{\Delta Q}{2f}\right)\sqrt{\frac{1-t}{1+t}}\;, \label{m}
\end{equation}
where
\begin{equation}
t=\sin\left(\frac{\pi\alpha_0}{2\phi}-\frac{\pi A\Delta\sigma}{16f\phi}\right)\;. \label{n}
\end{equation}
We can also determine the change in the corrective tractions in the permanent stick zone as
\begin{equation}
[q_{2}^{\ast}-q_{1}^{\ast}](x)=-\frac{4f \phi}{\pi A} \ln\left\vert \frac{1-\sqrt{(1-x/n)/(1+x/m)}}{1+\sqrt{(1-x/n)/(1+x/m)}} \right\vert \;,
\end{equation}
from (\ref{pressure_distribution_wedge}, \ref{mapping}).

\subsection{Display of example problem results}

\begin{figure}[H]
	\centering
	\includegraphics[scale=0.5, trim= 0 0 0 0, clip]{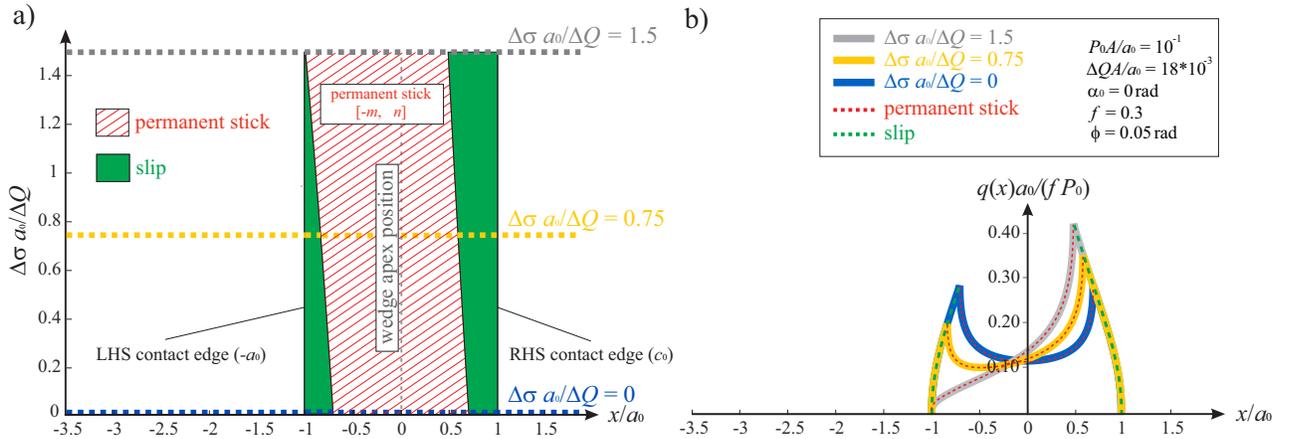}
	\caption{Steady state solutions of (a) slip and stick zone extents and (b) shear tractions for different values of a change in bulk stress, $\Delta\sigma$.}
	\label{fig:evolution_of_slipstick_with_DeltaSigma}	
\end{figure}

\hspace{0.4cm}
There is a multitude of possibilities to illustrate the results of the mapping procedure. The most sensible way seems to be to present, first, different steady state solutions for a varying bulk stress, $\Delta\sigma$. Consider Figure \ref{fig:evolution_of_slipstick_with_DeltaSigma} (a), note that each value of $\Delta\sigma a/\Delta Q$ represents a stand-alone steady state solution subject to a constant normal load, $P_0$, giving the mean contact size spanning $[-a_0, \quad c_0]$ and subject to a constant shear load fluctuation, $\Delta Q$, giving the permanent stick zone spanning $[-m, \quad n]$. In this first example the wedge is indented symmetrically, i.e. $\alpha_0$ is kept zero and the contact extent and position are unaffected by the change in bulk stress, ($a=c=$ const.). We see that, as the change in bulk stress is increased, the permanent stick zone shifts towards the LHS contact edge. Note however, the extent of the permanent stick zone increases slightly with a bigger $\Delta\sigma$, albeit almost not visible in this example. The solution finds its limit when the boundary of the permanent stick zone and a contact edge coincide, e.g. $a=m$ or $n=c$. If the change in bulk stress is increased further from that point onwards, the contact will experience reversed slip.   

In Figure \ref{fig:evolution_of_slipstick_with_DeltaSigma} (b), the shear traction distribution is depicted for three particular values of $\Delta\sigma a/\Delta Q$ as the slip zones reach their maximum extent in the steady state cycle and stick is only maintained in the permanent stick zone. Note that the \textit{area} underneath each shear traction distribution remains constant, i.e. $\int_{-a}^{c} q(x) \, \mathrm{d}x =$ const., and it is merely the form and slip-stick boundary that change with the steady state change in bulk stress.

\begin{figure}[H]
	\centering
	\includegraphics[scale=0.5, trim= 0 0 0 0, clip]{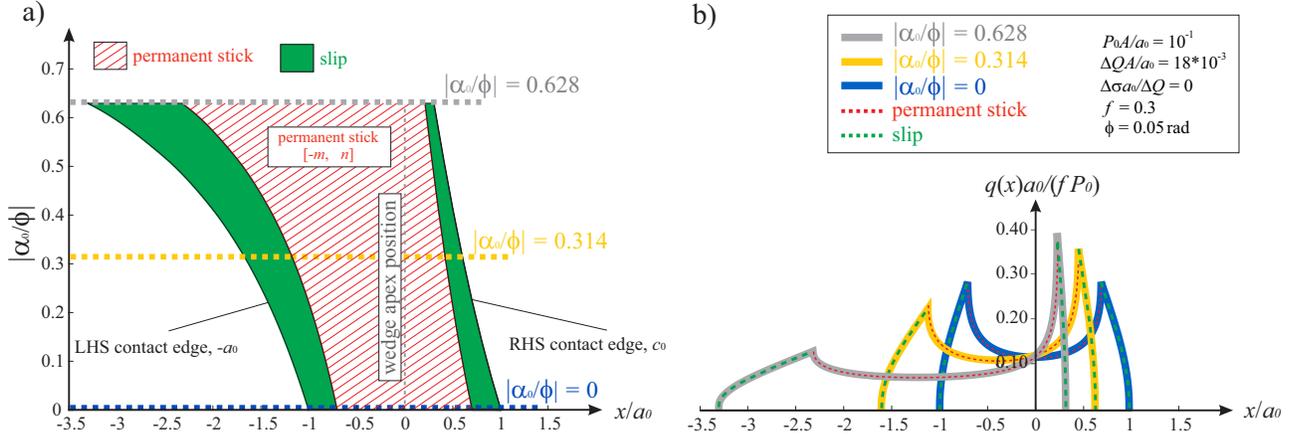}
	\caption{Steady state solutions of (a) slip and stick zone extents and (b) shear tractions for different values of a change in average angle of tilt, $\alpha_0$.}
	\label{fig:evolution_of_slipstick_with_alpha}	
\end{figure}

We now turn to the effects of the average angles of tilt, $\alpha_0$. Figure \ref{fig:evolution_of_slipstick_with_alpha} (a) is similar to Figure \ref{fig:evolution_of_slipstick_with_DeltaSigma} (a), each value of $\alpha_0$ represents a stand-alone steady state solution subject to a constant normal load, $P_0$, giving the mean contact size spanning $[-a_0, \quad c_0]$ and subject to a constant shear load fluctuation, $\Delta Q$, giving the permanent stick zone spanning $[-m, \quad n]$.  Here, the fluctuation in bulk stress, $\Delta\sigma$, is kept zero as the intention is to demonstrate the effects of tilt exclusively.

From Figure \ref{fig:evolution_of_slipstick_with_alpha} (a) it is apparent that the normal load solution dictates the qualitative behaviour of the tangential solution. Increasing the average angle of tilt influences not only the extent and position of the contact, but also the extent and position of the permanent stick zone within the contact. In Figure \ref{fig:evolution_of_slipstick_with_alpha} (b), the shear traction distribution is depicted for three particular values of $\alpha_0/\phi$ as the slip zones reach their maximum extent in the steady state cycle and stick is only maintained in the permanent stick zone. Note that, as in Figure \ref{fig:evolution_of_slipstick_with_DeltaSigma} (b), the \textit{area} underneath each shear traction distribution remains constant, i.e. $\int_{-a}^{c} q(x)\, \mathrm{d}x =$ const., and it is merely the extent, form, and slip-stick boundary that change with the average angle of tilt in the steady state.

In summary, the  two above figures reflect the solution behaviour indicated by equations \eqref{m} and \eqref{n}. We note that it is the mean normal load, $P_0$, and the average angle of tilt, $\alpha_0$, which affect the tangential solution, together with the tangential load inputs, $\Delta\sigma$ and $\Delta Q$.

\section{Conclusions}
\hspace{0.4cm}The paper provides a comprehensive but manageable method
for finding the size of the permanent stick zone for the problem of
a general half plane contact subject to a constant set of loads ($P,Q,M,\sigma$)
together with periodic changes in these same quantities.
 
The solution is appropriate when the bulk stress is never high enough to reverse
the sense of slip at a contact edge, and applies in the steady state. We found that the permanent stick zone is independent of the mean loads of shear $Q_0$ and bulk tension $\sigma_0$, but rather depends on $P_0$, $\Delta Q$, $\Delta\sigma$, and the average angle of tilt, $\alpha_0$. A straightforward way of using the full solution for normal load to find the size of the permanent stick zone is developed which requires
no further work or algebra to be carried out. The method is applied explicitly to the problem of a wedge as this permits all the algebraic steps needed to be displayed. Its application to the more frequent practically occurring ‘flat and rounded’ contact case is straightforward, if algebraically more taxing.

As we are interested in the maximum extent of the slip zones
during the steady state, ($a_{i}-m$) and ($c_{i}-n$), we need $\Delta P$ and $\Delta M$ as additional inputs for the normal load problem in order to obtain the contact coordinates at load points 1 and 2, $[-a_{i},\quad c_{i}]$, for $i=1,2$. However, these additional inputs and outputs are not required for carrying out the mapping procedure and finding the permanent stick zone.

\section*{Acknowledgements}
\hspace{0.4cm}This project has received funding from the European Union's Horizon 2020 research and innovation programme under the Marie Sklodowska-Curie agreement No 721865. David Hills thanks Rolls-Royce plc and the EPSRC for the support under the Prosperity Partnership Grant “Cornerstone: Mechanical Engineering Science to Enable Aero Propulsion Futures”, Grant Ref: EP/R004951/1.

\bibliography{Contribs_References_Hendrik_Jan2019}

\end{document}